\documentclass[conference]{IEEEtran}
\usepackage{blindtext, graphicx}
\usepackage{subcaption}
\usepackage{flushend}
\usepackage{float}

\ifCLASSINFOpdf
\else
\fi

\begin{document}

\title{Multi-limb Split Learning for Tumor Classification on Vertically Distributed Data }

 \author{\IEEEauthorblockN{Omar S. Ads\IEEEauthorrefmark{1},
 Mayar M. Alfares\IEEEauthorrefmark{1},
 Mohammed A.-M. Salem\IEEEauthorrefmark{1}\IEEEauthorrefmark{2}}
 \IEEEauthorblockA{\IEEEauthorrefmark{1}Faculty of Media Engineering \& Technology, German University in Cairo, Egypt}
 \IEEEauthorblockA{\IEEEauthorrefmark{2}Faculty of Computer and Information Sciences, Ain Shams University, Cairo, Egypt\\
 omar.ads@student.guc.edu.eg , mayar.mohamed@guc.edu.eg , mohammed.salem@guc.edu.eg}} 

\maketitle

\begin{abstract}
Brain tumors are one of the life-threatening forms of cancer. Previous studies have classified brain tumors using deep neural networks. In this paper, we perform the later task using a collaborative deep learning technique, more specifically split learning. Split learning allows  collaborative learning via neural networks splitting into two (or more) parts, a client-side network and a server-side network. The client-side is trained to a certain layer called the cut layer. Then, the rest of the training is resumed on the server-side network. Vertical distribution, a method for distributing data among organizations, was implemented where several hospitals hold different attributes of information for the same set of patients. To the best of our knowledge this paper will be the first paper to implement both split learning and vertical distribution for brain tumor classification. Using both techniques, we were able to achieve train and test accuracy greater than 90\% and 70\%, respectively.
\end{abstract}

\begin{IEEEkeywords}
Split Learning, Collaborative Learning, Federated Learning, Vertical Distribution, Brain Tumor Classification.
\end{IEEEkeywords}

\IEEEpeerreviewmaketitle

\section{Introduction}
Brain tumors are a mass or growth of abnormal cells in the brain and represent one of the deadliest forms of cancer. Adults who are diagnosed with severe types of brain tumors (i.e. glioblastoma) will most likely die after 2 years of the diagnosis \cite{website1}. Moreover, brain tumors are the most common cancer in children from 0 to 14 years old in the United States and children who survive it and enter adulthood will most likely be affected by the long-term as a consequences of all the surgeries, chemotherapy and radiotherapy \cite{article2}.

Therefore, deep learning for brain tumor classification has been done multiple times before by different approaches. However, the data required for training can not be gathered easily due to the health and privacy regulations, such as the GDPR (General Data Protection Regulation) \cite{article7} in Europe and the HIPAA (Health Insurance Portability and Accountability Act) in the USA which require the establishment of laws and standards to protect sensitive patient health information from disclosure without the consent or knowledge of the patient \cite{article19}.
These regulatory processes are crucial as medical data are highly sensitive and confidential. In addition, patients' medical records should not be shared or disclosed to any third party in any unauthorised way. 

The risk of data processing during deep learning training mainly depends on the context in which they are used. In contrast, to increase the accuracy of deep neural networks, collection of a large dataset is needed, while medical datasets tend to be relatively small in size.

Furthermore, Deep neural networks have a lot of parameters that requires huge amount of computational power which make it difficult for individual data repositories to train \cite{article16}. Distributed deep learning was introduced to solve these problems by distributing the network over multiple organizations (i.e. hospitals), thus, having more computational power while preserving privacy without sharing the patients' raw data.

In this paper, we present an approach to classify vertically distributed brain MRI images into either healthy or tumorous using split learning.

The paper is organized as follows: a literature review is presented in section II, the methodology and implementation are written in section III. The results are analyzed in section IV. Finally, the conclusion and the future work are discussed in Section V.

\section{Background}
Brain tumor classification using deep learning techniques is a popular topic and has been employed multiple times with different approaches. However, at the best of our knowledge, the task was never performed in a split learning setup. 

Privacy-preserving AI was introduced to address the growing demand of machine learning applications which comes with the question of privacy and how to protect the user's data.

In this section, we give an introduction about the different topics used in this study and discuss the related previous works.

\subsection{Brain tumor classification with Convolutional Neural Networks} 
Deep learning mimics the working of the human brain in the way of processing data and creating patterns for decision making. 
Usually, when it comes to image classification in deep neural networks the most common approach is Convolutional Neural Networks (CNN). CNN is a type of neural network that specializes in processing data with a grid-like topology, such as images \cite{article23}. 

In a study done by Diaz-Pernass et al. \cite{article14}, CNNs were used to train a model on brain tumor classification. Three types of tumors were classified: meningioma, glioma, and pituitary tumors. The dataset consisted of 3064 MRI images from 233 patients and were able to achieve accuracy of 97.3 \% due to three spatial scales along different processing pathways. 

In another study done by Das et al. \cite{article20}, CNNs were used for classifying brain tumors in T1-weighted contrast-enhanced MRI images. The system consisted of two parts: image processing followed by image classification. The dataset consists of 3064 images that contain 3 types of tumors: glioma, meningioma and pituitary and was able to achieve a testing accuracy of 94.39\%, average precision of 93.33\% and an average recall of 93\%.

Saxena et al. \cite{article25} implemented 3 pre-trained CNN models to classify MRI images into 2 classes. A dataset of 253 images was split into 183, 50 and 20 for training, validation and testing respectively. The best performing model was a Resnet-50 model with accuracy of 90\%, in the second place came VGG-16 with accuracy of 90\% and finally the last one was Inception-V3 model with a 55\% accuracy.

Finally, in a study by Sultan et al. \cite{article26}, a CNN-based model was used to train on two brain MRI datasets of size 3064 and 516, respectively. They were able to achieve an accuracy of  96.13\% and 98.7\% for the first and second datasets, respectively.

CNNs represent many challenges such as overfitting, exploding gradients, class imbalance and the need of large datasets. They require high computational power and long time to train which introduce lots of research limitations \cite{book1}. Split learning was developed to tackle these issues where the computational power is distributed among organisations. It also has the advantage of keeping the data secure and private unlike conventional deep neural networks.

\subsection{Privacy-Preserving Machine Learning (PPML)}

Deep  Learning  (DL)  systems  have  achieved  noticeable  performance and  are  being extensively used in different domains.  The recent privacy violation incidents and dataleakage incited the need for developing privacy-preserving approaches inside DL systems to be able to preserve data confidentiality and user privacy. PPML is a DL technique in which user privacy and data security are preserved. In Westin \cite{citeKeyMayar},  information privacy was defined as ”the claim of individuals,  groups,  orinstitutions to determine for themselves when, how, and to what extent information about them is communicated to others.

\subsection{Distributed deep learning (DDL)}

In conventional deep learning, the user data is sent to a centralized server which involves privacy risks like data leakage and possible misuse. DDL was mainly introduced to solve this issue. DDL was further modified to address the computational power problem of conventional deep learning via multiple GPUs \cite{article5}. Thus, in this paper, referring to DDL includes both Federated learning and Split learning.

\subsubsection{Federated Learning (FL)}
FL is a type of collaborative deep learning where each device sends its own copy of the model to be trained on each device on its corresponding local data. Then, the updated model parameters are sent back to the server using encrypted communication. The updates to the model are sent to the cloud to be averaged in order to improve the shared model. This is done to protect the users data since data is not shared with the server, only the updated model parameters are exchanged \cite{article6}. Therefore, FL allows access to multiple resources from different organizations (i.e. hospitals) which increases the computational power. However, in the real world, organizations have low computing resources so training large models locally is not feasible \cite{article1}. Thus, Split Learning is employed.

\subsubsection{Split Learning (SL)}
\begin{figure}[H]
\centering
\includegraphics[width= 8.7cm]{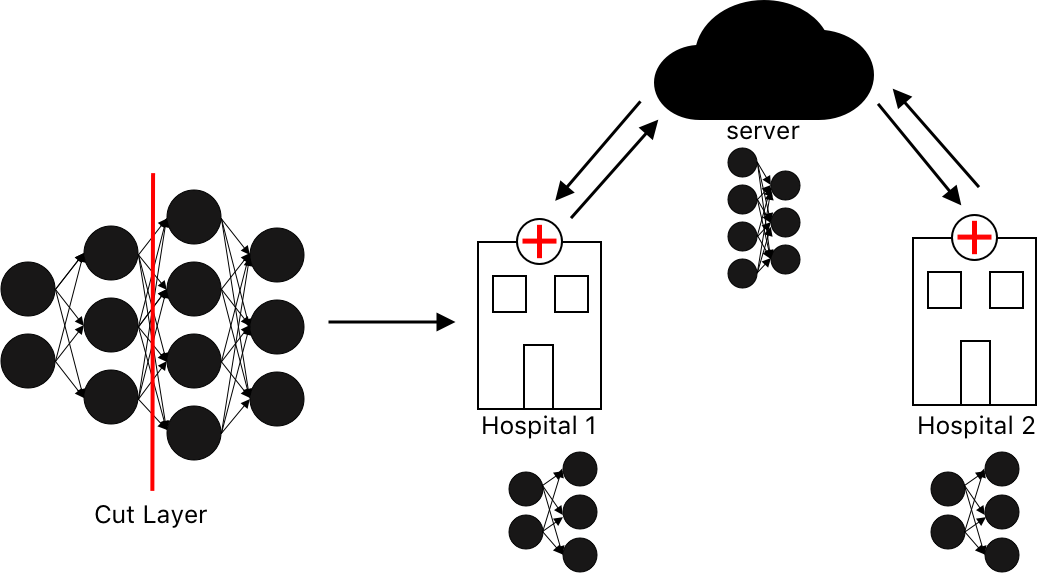}
\caption{Cutting of the neural network}
\label{figure:split}
\end{figure}

\begin{figure}[H]
\centering
\begin{subfigure}{0.35\textwidth}
\centering
\includegraphics[width= 1\textwidth]{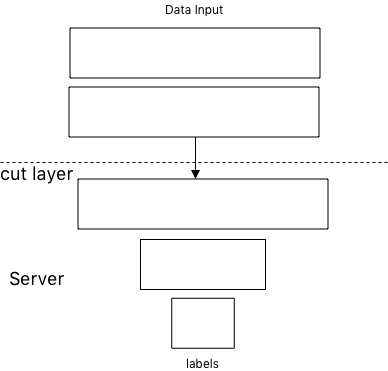}
\caption{Simple vanilla data distribution}
\label{figure:simple}
\end{subfigure}
\begin{subfigure}[h]{0.45\textwidth}
  \centering
  \includegraphics[width=1\textwidth]{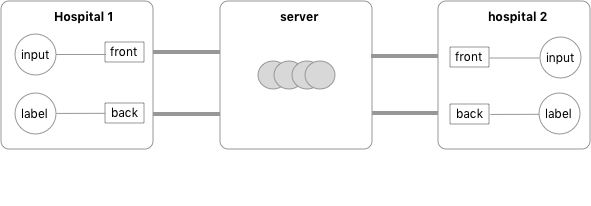}
  \caption{U-shaped data distribution}
  \label{figure:u-shaped}
\end{subfigure}

\begin{subfigure}[h]{0.35\textwidth}
  \centering
  \includegraphics[width=1\textwidth]{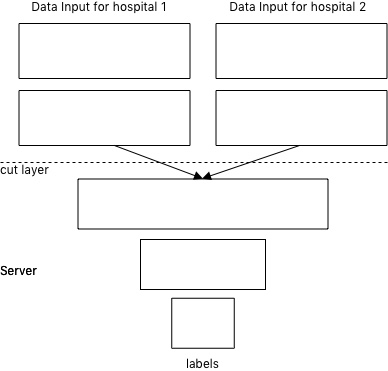}
  \caption{Vertically distributing data among organizations}
  \label{figure:dis}
\end{subfigure}
\caption{Data distribution in Split Learning}
\end{figure}

SL is a type of distributed deep learning where the network is split into two parts: a client-side (CS) and a server-side (SS) as illustrated in figure \ref{figure:split}. The client side can be formed of multiple clients. Those clients start the training until a certain layer known as the cut layer. The activation of the cut layer, called the smashed data, is sent to the server side. The server side continues the training on the rest of the layers. This represents a single forward propagation. Then, the server side carries out the back propagation up to the cut layer and sends the gradients of the smashed data back to the clients side. This represents a single backpropagation between the clients and the server. The process is repeated until convergence or until a certain number of rounds is reached \cite{article10,article1}.
\newline In split learning, the architecture and the dimension of the input feature vector at each layer determines the cost of the computation \cite{article9}.
\newline Data distribution in split learning can have many forms (i.e. U-shaped, Simple vanilla and Vertical distribution). Simple vanilla is the simplest type of distributing data where the client (i.e. hospital) trains the network up to a specified cut layer and then send the activation of the cut layer to the sever to continue the rest of the training as illustrated in figure \ref{figure:simple}. Unlike vertical distribution (explained in subsection D) and Simple vanilla where the labels are shared with the server,  U-shaped data distribution is applied when there is no need to share the labels with the server and this is done when labels are highly sensitive. In this configuration, the neural network were folded into the end layers of the server network and the output was sent back to the client entity while most of the server layers are still retained. Clients generate the gradients from the end layers and use them for backpropagation without sharing the labels as illustrated in figure \ref{figure:u-shaped} \cite{article16,article10,article13}. 
Split learning introduces a new kind of parallelism which is parallelization among parts of a model. It also provides a huge reduction to the computational burden at each participant while maintaining the model performance. Gupta et al.\cite{article24} found that split learning reduces leakage thus improving security and decreases the transmitted data thus improving communication too.

In a study done by Vepakomma et al. \cite{article13}, split learning was tested on a CIFAR-10 and CIFAR-100 datasets using VGG and Resnet-50 architectures for 100 and 500 client-based setups respectively. They achieved 0.1548 and 0.03 TFlops in terms of computation resources, and 6 and 1.2 GB in terms of computation bandwidth. Results showed that split learning outperformed both federated learning and conventional deep leaning.

In another study done by Poirot et al. \cite{article10}, split learning on a 156,535 chest X-rays dataset with different collaborative setups ranging from 1-50 clients using u-shaped data distribution achieved a split mean of 0.888, 0.850 and 0.859 for the 1, 25 and 50 clients respectively. Results showed that distributed learning settings can give a great performance boost in comparison to non-collaborative settings.

\subsection{Vertically Distributed Data}
Vertically distributed data is achieved when multiple entities (i.e. hospitals) hold various aspects of information from the same set of participants (i.e. patients) as illustrated in \ref{figure:dis}. Vertical distribution can be considered as a unique case of arbitrary distribution where data related to the same set of entities is included in all partitions. This allows different entities holding different modalities of data to train distributed models without sharing their raw data \cite{article11,article12}. In addition, vertical distribution is the most secure and private data distribution \cite{article13}, thus, it is used in our study. To the best of our knowledge, vertically distributed datasets are not yet available so we synthesized the distribution in our work as discussed in section III-B.

\section{Methodology}
\label{Methodology}
In this work, we mimicked a real world situation with two entities, namely Hospital 1 and Hospital 2 and a server. Each hospital's data is siloed. Moreover, the server does not have access to the raw data (i.e. medical scans), only the labels are sent to the server. Each hospital starts the training on its local data. Then, a cut layer is defined and the smashed data is sent to the server to resume training.
In other words, Hospital 1 cannot access the data of Hospital 2 and vise versa. The server cannot see the raw data nor access it and is given only the labels of the MRI images.

OpenMined Duet \cite{duet} was used in our implementation in order to move away from the central server management ideology. Duet allows distributed learning setups to form a network of organisations to communicate with each other and train a model without sharing their data while keeping the network secure and private. 

\subsection{Dataset}
MRI is the most widely used type of medical imaging due to its image quality and its dependency on non-ionizing radiation. Physicians face a challenge when studying MRI images due to the time and effort they have to put in diagnosing the brain tumor \cite{article3}.
\newline In this paper, we used a publicly available dataset of MRI images from Kaggle which was gathered from multiple resources and contributions \cite{dataset}.
The dataset consists of 4600 MRI images of the human brain, but only 2000 images were used in our work due to computational power restrictions. Images are classified into healthy and tumorous brains. The dataset is splitted into 1600 and 400 images for training and testing, respectively.

\subsection{Data Pre-Processing and Vertical Distribution}

\begin{figure}[H]
\centering
\begin{subfigure}{0.26\textwidth}
\centering
\includegraphics[width= 0.9\textwidth]{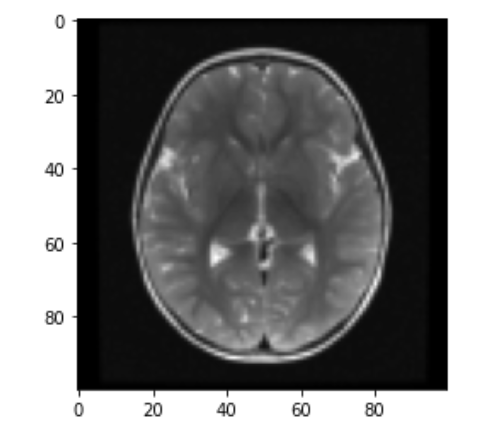}
\caption{Image before vertical distribution}
\label{figure:brain}
\end{subfigure}
\begin{subfigure}[ht]{0.23\textwidth}
  \centering
  \includegraphics[width=0.65\textwidth]{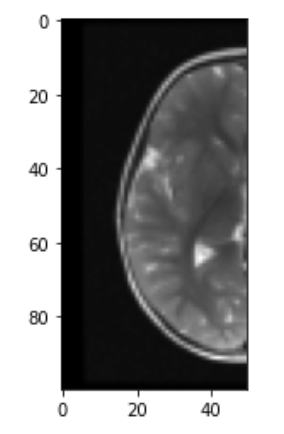}
  \caption{Image given to hospital 1}
  \label{figure:brain1}
\end{subfigure}

\begin{subfigure}[ht]{0.23\textwidth}
  \centering
  \includegraphics[width=0.65\textwidth]{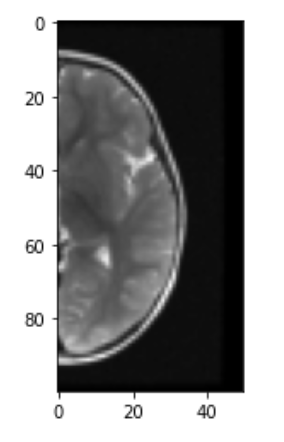}
  \caption{Image given to hospital 2}
  \label{figure:brain2}
\end{subfigure}
\caption{vertically distributing the MRI image}
\label{figure:brain3}
\end{figure}

Images are first resized to 100x100 to better fit our model then transformed to grayscale.
Moreover, the MRI images were vertically distributed among both hospitals.
Then, each MRI was split into two halves and each half was given to one of the hospitals. The input and outputs are illustrated in figures \ref{figure:brain}, \ref{figure:brain1} and \ref{figure:brain2}, respectively, where image \ref{figure:brain1} is given to the first hospital and image \ref{figure:brain2} is given to the second hospital \cite{article8}. This was done to mimic the real world situation; for example, a hospital may have the records of the patient and a radiology center hold the MRI of the same patient both train on their own records up to a cut layer and then send the smashed data to a cancer classifier server to continue the rest of the training. This offers protection to the patients data and is faster than a federated learning model where all of the training is done locally. Unfortunately, due to the lack of datasets having the above-mentioned criteria, we mimicked the same situation by giving different parts of the same image to both hospitals as a proof-of-concept.

\subsection{Model}
The model consists of 2 parts: The first one is on the hospitals side where each hospital have its own model consisting of 1 layer followed by Relu activation function, to reduce the computational power at the hospital side. The smashed data is then sent to the server to be trained on 2 additional layers followed by a Relu activation function. 
Relu was mainly used in order to reduce the model complexity. It is a simple function with a small computation time. Thus, the execution is faster and the overall training time is decreased. Furthermore, Relu allows the network to learn the complex patterns in the data \cite{article17}. 
\newline We also used SDG (Stochastic Gradient Descent) in our network on both hospitals and the server. SDG perform one update at a time. Therefore, it is usually much faster and can also be used for collaborative learning \cite{article18}.

\subsection{Training and Testing}
First of all, hospitals 1 and 2 start training until the specified cut layer is reached. Then, both hospitals send the activation of the cut layer (the smashed data) to the server to finish a single forward propagation. Afterwards, the server carries out the back propagation up to the cut layer, sends the gradients of the smashed data back to the hospitals' side and finishes a single back propagation. This loop will continue until the model converges or a certain number of epochs is reached. Finally, the training and test accuracies are calculated.

\section{Results}
\label{Results}
\begin{figure}[H]
\centering
\includegraphics[width= 0.41\textwidth]{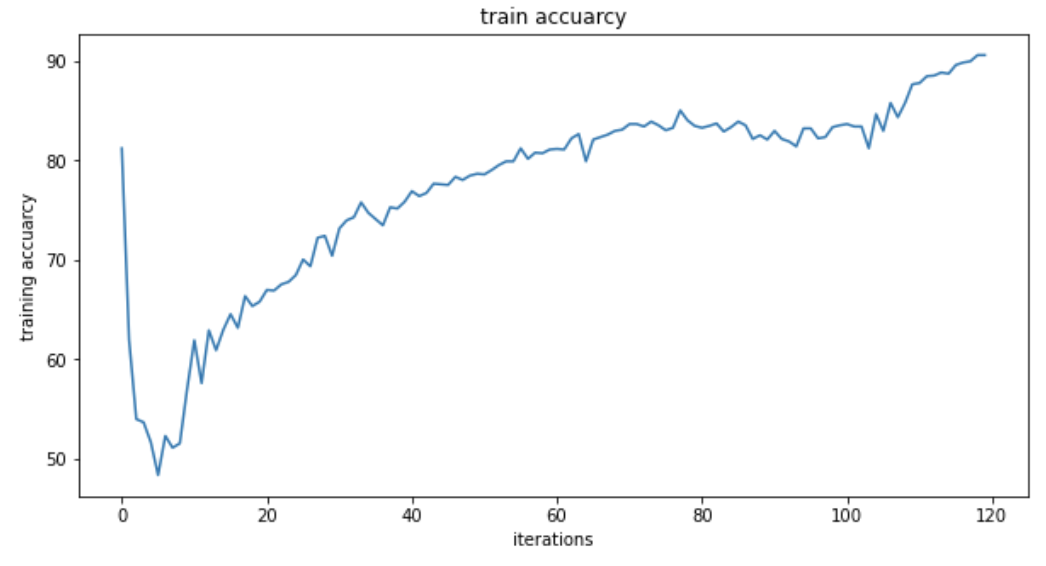}
\caption{Training Accuracy}
\label{figure:train}
\end{figure}

\begin{figure}[H]
\centering
\includegraphics[width= 0.41\textwidth]{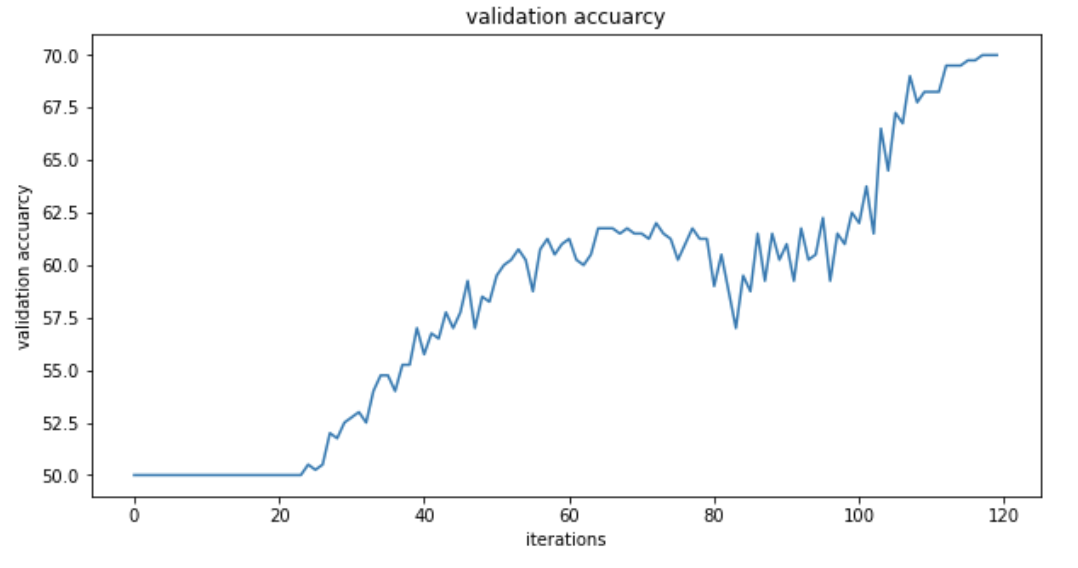}
\caption{Test Accuracy}
\label{figure:validation}
\end{figure}

We were able to achieve a training accuracy of 90.688\% after training on 120 epoch as shown in figure \ref{figure:train}.  Meanwhile , we reached a test accuracy of 70.250\% as seen in figure \ref{figure:validation}. The results of the test accuracy could be enhanced if trained on a lager number of epochs but due to limited computational power, between our local GPUs and the Duet servers restrictions, we were not able to conduct further training.

\section{Conclusion and Future Work}
Distributed deep learning is needed in the medical scope in order to advance the treatment and diagnosis of diseases, especially cancer. As proven in previous studies by Poirot et al. \cite{article16} and Vepakomma et al. \cite{article18}, split learning outperforms most collaborative learning techniques while reducing the leakage of information. This could be an incentive for healthcare institutions to implement split learning for future advancement. SL provides privacy, distributed computational power and less infrastructure modifications with a reduced cost.

Split learning is a relatively new topic and frameworks that support it are still in the developing phase. In addition, datasets supporting vertical distribution are not publicly available and require a lot of resources. Thus, more research is required.

\ifCLASSOPTIONcaptionsoff
  \newpage
\fi

\bibliographystyle{IEEEtran}
\bibliography{citations}

% Generated by IEEEtran.bst, version: 1.14 (2015/08/26)
\begin{thebibliography}{10}
\providecommand{\url}[1]{#1}
\csname url@samestyle\endcsname
\providecommand{\newblock}{\relax}
\providecommand{\bibinfo}[2]{#2}
\providecommand{\BIBentrySTDinterwordspacing}{\spaceskip=0pt\relax}
\providecommand{\BIBentryALTinterwordstretchfactor}{4}
\providecommand{\BIBentryALTinterwordspacing}{\spaceskip=\fontdimen2\font plus
\BIBentryALTinterwordstretchfactor\fontdimen3\font minus
  \fontdimen4\font\relax}
\providecommand{\BIBforeignlanguage}[2]{{%
\expandafter\ifx\csname l@#1\endcsname\relax
\typeout{** WARNING: IEEEtran.bst: No hyphenation pattern has been}%
\typeout{** loaded for the language `#1'. Using the pattern for}%
\typeout{** the default language instead.}%
\else
\language=\csname l@#1\endcsname
\fi
#2}}
\providecommand{\BIBdecl}{\relax}
\BIBdecl

\bibitem{website1}
\BIBentryALTinterwordspacing
K.~Aldape, K.~M. Brindle, and L.~Chesler. (2019) Challenges to curing primary
  brain tumours. [Online]. Available:
  \url{https://doi.org/10.1038/s41571-019-0177-5}
\BIBentrySTDinterwordspacing

\bibitem{article2}
T.~M. Brinkman, M.~J. Krasin, W.~Liu, G.~T. Armstrong, R.~P. Ojha, Z.~S.
  Sadighi, P.~Gupta, C.~Kimberg, D.~Srivastava, T.~E. Merchant, A.~Gajjar,
  L.~L. Robison, M.~M. Hudson, and K.~R. Krull, ``Long-term neurocognitive
  functioning and social attainment in adult survivors of pediatric cns tumors:
  Results from the st jude lifetime cohort study,'' \emph{Journal of Clinical
  Oncology}, vol.~34, no.~12, pp. 1358--1367, 2016.

\bibitem{article7}
\BIBentryALTinterwordspacing
N.~B. Truong, K.~Sun, S.~Wang, F.~Guitton, and Y.~Guo, ``Privacy preservation
  in federated learning: Insights from the {GDPR} perspective,'' \emph{CoRR},
  vol. abs/2011.05411, 2020. [Online]. Available:
  \url{https://arxiv.org/abs/2011.05411}
\BIBentrySTDinterwordspacing

\bibitem{article19}
G.~J. Annas, ``Hipaa regulations—a new era of medical-record privacy?'' 2003.

\bibitem{article16}
\BIBentryALTinterwordspacing
O.~Gupta and R.~Raskar, ``Distributed learning of deep neural network over
  multiple agents,'' \emph{Journal of Network and Computer Applications}, vol.
  116, pp. 1--8, 2018. [Online]. Available:
  \url{https://www.sciencedirect.com/science/article/pii/S1084804518301590}
\BIBentrySTDinterwordspacing

\bibitem{article23}
J.~Schmidhuber, \emph{Deep learning in neural networks: An overview}.\hskip 1em
  plus 0.5em minus 0.4em\relax Elsevier, 2015, vol.~61.

\bibitem{article14}
F.~J. D{\'\i}az-Pernas, M.~Mart{\'\i}nez-Zarzuela, M.~Ant{\'o}n-Rodr{\'\i}guez,
  and D.~Gonz{\'a}lez-Ortega, ``A deep learning approach for brain tumor
  classification and segmentation using a multiscale convolutional neural
  network,'' in \emph{Healthcare}, vol.~9, no.~2.\hskip 1em plus 0.5em minus
  0.4em\relax Multidisciplinary Digital Publishing Institute, 2021, p. 153.

\bibitem{article20}
S.~Das, O.~R.~R. Aranya, and N.~N. Labiba, ``Brain tumor classification using
  convolutional neural network,'' pp. 1--5, 2019.

\bibitem{article25}
P.~Saxena, A.~Maheshwari, and S.~Maheshwari, ``Predictive modeling of brain
  tumor: A deep learning approach,'' in \emph{Innovations in Computational
  Intelligence and Computer Vision}, M.~K. Sharma, V.~S. Dhaka, T.~Perumal,
  N.~Dey, and J.~M. R.~S. Tavares, Eds.\hskip 1em plus 0.5em minus 0.4em\relax
  Singapore: Springer Singapore, 2021, pp. 275--285.

\bibitem{article26}
H.~H. Sultan, N.~M. Salem, and W.~Al-Atabany, ``Multi-classification of brain
  tumor images using deep neural network,'' \emph{IEEE Access}, vol.~7, pp.
  69\,215--69\,225, 2019.

\bibitem{book1}
S.~Joshi, D.~Verma, G.~Saxena, and A.~Paraye, \emph{Issues in Training a
  Convolutional Neural Network Model for Image Classification}, 07 2019, pp.
  282--293.

\bibitem{citeKeyMayar}
A.~F. Westin, \emph{{Privacy And Freedom}}.\hskip 1em plus 0.5em minus
  0.4em\relax 25 Wash. \& Lee L. Rev. 166, 1968.

\bibitem{article5}
\BIBentryALTinterwordspacing
M.~Langer, Z.~He, W.~Rahayu, and Y.~Xue, ``Distributed training of deep
  learning models: A taxonomic perspective,'' \emph{IEEE Transactions on
  Parallel and Distributed Systems}, vol.~31, no.~12, p. 2802–2818, Dec 2020.
  [Online]. Available: \url{http://dx.doi.org/10.1109/TPDS.2020.3003307}
\BIBentrySTDinterwordspacing

\bibitem{article6}
H.~B. McMahan, E.~Moore, D.~Ramage, S.~Hampson, and B.~A. y~Arcas,
  ``Communication-efficient learning of deep networks from decentralized
  data,'' 2017.

\bibitem{article1}
C.~Thapa, M.~A.~P. Chamikara, and S.~Camtepe, ``Splitfed: When federated
  learning meets split learning,'' 2020.

\bibitem{article10}
M.~G. Poirot, P.~Vepakomma, K.~Chang, J.~Kalpathy-Cramer, R.~Gupta, and
  R.~Raskar, ``Split learning for collaborative deep learning in healthcare,''
  2019.

\bibitem{article9}
I.~Ceballos, V.~Sharma, E.~Mugica, A.~Singh, A.~Roman, P.~Vepakomma, and
  R.~Raskar, ``Splitnn-driven vertical partitioning,'' \emph{arXiv preprint
  arXiv:2008.04137}, 2020.

\bibitem{article13}
P.~Vepakomma, O.~Gupta, T.~Swedish, and R.~Raskar, ``Split learning for health:
  Distributed deep learning without sharing raw patient data,'' 2018.

\bibitem{article24}
O.~Gupta and R.~Raskar, ``Distributed learning of deep neural network over
  multiple agents,'' 2018.

\bibitem{article11}
S.~Hardy, W.~Henecka, H.~Ivey-Law, R.~Nock, G.~Patrini, G.~Smith, and
  B.~Thorne, ``Private federated learning on vertically partitioned data via
  entity resolution and additively homomorphic encryption,'' 2017.

\bibitem{article12}
S.~Agrawal, V.~Narasayya, and B.~Yang, ``Integrating vertical and horizontal
  partitioning into automated physical database design,'' in \emph{Proceedings
  of the 2004 ACM SIGMOD international conference on Management of data}, 2004,
  pp. 359--370.

\bibitem{duet}
\BIBentryALTinterwordspacing
T.~Jain, ``Duet demo - how to do data science on data owned by a different
  organization,'' Jan 2021. [Online]. Available:
  \url{https://blog.openmined.org/duet-demo-how-to-do-data-science-on-data-owned-by-a-different-organization/}
\BIBentrySTDinterwordspacing

\bibitem{article3}
N.~B. Bahadure, A.~K. Ray, and H.~P. Thethi, ``Image analysis for mri based
  brain tumor detection and feature extraction using biologically inspired bwt
  and svm,'' \emph{International journal of biomedical imaging}, vol. 2017,
  2017.

\bibitem{dataset}
\BIBentryALTinterwordspacing
P.~Viradiya, ``Brian tumor dataset,'' May 2021. [Online]. Available:
  \url{https://www.kaggle.com/preetviradiya/brian-tumor-dataset}
\BIBentrySTDinterwordspacing

\bibitem{article8}
J.~Vaidya, \emph{Vertically Partitioned Data}.\hskip 1em plus 0.5em minus
  0.4em\relax Boston, MA: Springer US, 2009, pp. 3263--3265.

\bibitem{article17}
C.~Nwankpa, W.~Ijomah, A.~Gachagan, and S.~Marshall, ``Activation functions:
  Comparison of trends in practice and research for deep learning,''
  \emph{arXiv preprint arXiv:1811.03378}, 2018.

\bibitem{article18}
S.~Ruder, ``An overview of gradient descent optimization algorithms,'' 2017.

\end{thebibliography}

\vfill

\end{document}